# Integrating On-chain and Off-chain Governance for Supply Chain Transparency and Integrity

Shoufeng Cao
*QUT Design Lab*
Queensland University of Technology
Brisbane, Australia
shoufeng.cao@qut.edu.au

Thomas Miller
*QUT Design Lab*
Queensland University of Technology
Brisbane, Australia
tj.miller@hdr.qut.edu.au

Marcus Foth
*QUT Design Lab*
Queensland University of Technology
Brisbane, Australia
m.foth@qut.edu.au

Warwick Powell
*BeefLedger Ltd & Smart Trade Networks Ltd*
Brisbane, Australia
wp@smarttradenetworks.com

Xavier Boyen
*School of Computer Science*
Queensland University of Technology
Brisbane, Australia
xavier.boyen@qut.edu.au

Charles Turner-Morris
*BeefLedger Ltd & Smart Trade Networks Ltd*
Brisbane, Australia
cm@beefledger.io

*Abstract*—Integrating on-chain and off-chain data storage for decentralised and distributed information systems, such as blockchain, presents specific challenges for providing transparency of data governance and ensuring data integrity through stakeholder engagement. Current research on blockchain-based supply chains focuses on using on-chain governance rules developed for cryptocurrency blockchains to store some critical data points without designing tailored on-chain governance mechanisms and disclosing off-chain decision-making processes on data governance. In response to this research gap, this paper presents an integrated data governance framework that coordinates supply chain stakeholders with inter-linked on-chain and off-chain governance to disclose on-chain and off-chain rules and decision-making processes for supply chain transparency and integrity. We present a Proof-of-Concept (PoC) of our integrated data governance approach and suggest future research to strengthen scaling up and supply chain-based use cases based on our learnings.

*Keywords— Blockchain, on-chain, off-chain, governance, supply chain, data transparency, data integrity*

## I. INTRODUCTION

Blockchain is a decentralised and distributed ledger, which highlights the necessity of achieving consensus from various stakeholders on how to record and store transaction information on the shared ledger [1], [2]. This requires a proper governance mechanism to coordinate stakeholders with data injection, storage, sharing and access. Blockchain-based systems are generally governed by a set of rules to achieve a certain level of accountability, fairness, transparency, assurance, and stakeholder engagement. While there is a lack of commonly accepted definition associated with blockchain governance [3], De Filippi and Mcmullen [4] defined blockchain governance as either the governance of a blockchain system by infrastructure or the governance of a blockchain infrastructure system. They describe blockchain governance as either 'on-chain' or 'off-chain' patterns [4]. On-chain governance enables decision-making processes to be finalised within the blockchain system, which they consider to be formal, fair, and transparent, whereas off-chain governance finalises decision-making away from the blockchain system and this is considered informal and hidden most of the time [3].

Blockchain-enabled supply chains are one of the promising areas of blockchain application beyond the financial sector [5]. While various proof-of-concept work and use cases have been reported to develop blockchain-based supply chains, especially for supply chain traceability, current research focuses on using on-chain governance rules developed for cryptocurrency blockchains to store critical data points [6]. The application of blockchain to a multi-stakeholder supply chain can generate a large amount of data and accordingly increase the cost of data storage [7]. This creates demand for more cost-effective off-chain data storage. However, a proper disclosure of off-chain decision-making processes is often still lacking, which can jeopardise data transparency, compliance and integrity expectations. This paper reports ongoing research, which seeks to address this research gap by developing an integrated data governance framework that coordinates stakeholders with inter-linked on-chain and off-chain supply chain data. It enables users to disclose on-chain and off-chain rules and decision-making processes to the wider community via the Smart Trade Network (STN), which is a proof of authority Ethereum Virtual Machine compatible blockchain network.

Following this introduction section, we review related work on blockchain governance, and data transparency and integrity (II). This is followed by the description of our work-in-progress of an integrated data governance framework (III) before presenting the pilot use case implementation and evaluation results. We conclude with our learnings from our pilot use case and suggest future research areas.

## II. RELATED LITERATURE

### A. Blockchain Governance

Blockchain governance is a novel and complex issue as it deals with the coordination between human involvement and the automation of decision-making processes [4], [8]. The discussion of governance mechanisms in blockchain systems generally distinguishes between on-chain and off-chain governance [3]. Whether implementing on-chain or off-chain models, both typically employ mechanisms to make consensual decisions on data updates and validation. Various on-chain consensus mechanisms with a type of "proof," e.g., "proof of work," "proof of stake," "proof-of-authority" and other proofs have been developed for blockchain systems [9], [10]. The literature has emphasised the advantages of on-chain governance. However, they were largely developed to safeguard against (e.g.) Byzantine Fault Tolerance and Sybil Attacks [11] and each consensus mechanism is more suitable in a specific circumstance and less suitable in others due their own sets of risks and drawbacks [12]. Obviously, current on-chain governance mechanisms are not perfect for organising and safeguarding each stage of the supply chain against errors, fraud, and hardware failure in multi-stakeholder supply chains [9], [13]. Furthermore, as noted by De Filippi and Mcmullen [8], on-chain governance mechanisms have formal and rigid coding structures that restrict their flexibility to enable the system to react to unforeseen circumstances quickly and smoothly. Differently, off-chain governance has relatively informal and unstructured formats [8] and can complement the weakness of on-chain governance. Bitcoin and Ethereum are two typical examples that use a combination of offline coordination and online code modifications to implement update changes. However, they are criticised for allowing miners and developers to play the role in coordinating and achieving consensus between stakeholders in off-chain governance [14]. Therefore, an integrated on-chain and off-chain governance mechanism needs to be developed with stakeholders in ways that not only maintain the transparency and efficiency of on-chain governance rules, but also have the capability to balance the different needs and interests of stakeholders. Our study aims to develop and test an integrated blockchain governance model to coordinate on-chain and off-chain mechanisms for blockchain-based supply chains.

### B. Transparency and Integrity

Transparency and integrity are two key concepts related to trust [15], [16] in multi-tier supply chains that involve interdependent actors from production to consumption. Supply chain transparency refers to the disclosure of information to supply chain stakeholders about an agent's operations and how products are distributed to consumers [17]. Differently, supply chain integrity refers to adherence to good practice of being honest in supply chain activities and showing a consistent dedication to maintaining integrity in supply chain processes and flows [18]. Though they are different concepts, supply chain transparency is regarded as a prerequisite to supply chain integrity. This is supported by Stohl et al. [19] who argued that the transparency of decision processes and behaviours to others could force actors and companies to behave with integrity because of accountability. A transparent supply chain requires information disclosure to supply chain stakeholders and therefore prompts the collection and sharing of relevant information. At root, it is about reducing the extent of information asymmetry [2], [13]. Technology plays a role in addressing these issues. However, companies are worried about their private and confidential information [17]. Blockchain technology has its transparent, immutable, secure, and decentralised attributes and has the potential to address these issues. Recent research has explored the role of blockchain technology on supply chain transparency. Sunny et al. [20] found that better supply chain transparency can be achieved through blockchain-based traceability. Rao et al. [21] reported that blockchain plays an important role in improving supply chain transparency from two serialised data projects. Blockchain can also improve the integrity of data with immutable ledgers that effectively prohibits the alteration or deletion of recorded information. This allows data users to verify data integrity and authenticity. However, blockchain does not safeguard against intentional efforts to upload false and corrupted data, nor does it guarantee against unintentional error [13], [20]. Therefore, there is a need for transparency in decision-making processes on data entry and validation to ensure supply chain integrity, which is the focus of this paper.

## III. AN INTEGRATED GOVERNANCE FRAMEWORK

This paper proposes an integrated approach that enables the interaction between on-chain and off-governance, which leads to an integrated data governance framework as shown in Figure 1. The framework not only provides a way to ensure the transparency and efficiency of off-chain governance rules but also supports the achievement of flexibility in on-chain and off-chain governance. Both on-chain and off-chain governance are achieved through multi-sig smart contracts deployed on the STN network that can interact with the Ethereum Mainnet and Polygon/Matic network. The Polygon/Matic network is a scaling blockchain framework for developing Ethereum-compatible blockchain networks [22]. The off-chain governance leverages the smart contract created on the Polygon/Matic network to ensure the transparency of consensus-based decision-making in off-chain governance so as to improve supply chain transparency and integrity. Although these smart contracts could also be deployed on the STN network itself, we chose to deploy these off-chain governance support functions on the Matic network to explore its bridging functions to the Ethereum mainnet, enabling users to move tokens from one to the other, as well as the impact of variable transaction costs when the multisig governance contracts are deployed and executed. The STN network, by way of contract, does not have 'variable' fees per se, which makes it more suitable for stable cost-based business case development for industry application.

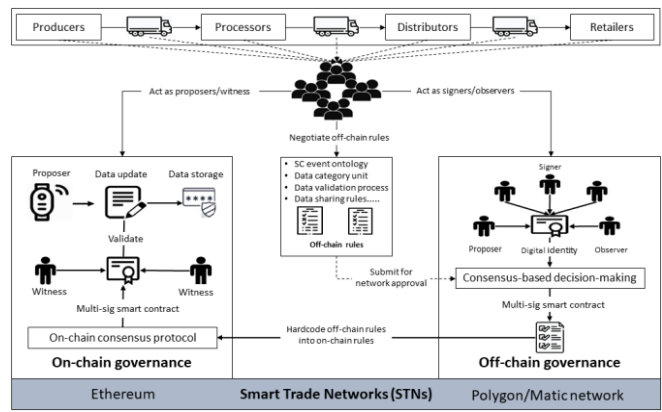

Fig. 1. An integrated data governance architecture for blockchain-based supply chains

### A. On-chain Governance Mechanism

The on-chain governance approach deployed on the STN network uses the primary multisig protocol built on top of the Ethereum Virtual Machine compatible blockchain. Multisig wallets are implemented as a smart contract to validate the data state updates and storage when data points are proposed on the network for validation following the pre-defined multisig protocol. The on-chain consensus protocol implemented by the multisig smart contract defines the rules, including the signatory group, the number of signatories and the process of validation, for writing data to the blockchain and validating data in the context of a multi-stakeholder supply chain [2]. The on-chain consensus protocol also defines the rights and obligations of each group member in creating and joining a signatory group to propose and accept/reject changes to the on-chain consensus protocol. The on-chain consensus protocol is being resolved using multisig "voting" on the Polygon/Matic network with a dedicated multisig mechanism.

## B. Off-chain Governance Mechanism

The off-chain governance is also deployed on the STN network but using secondary multi-sig protocols on the Polygon/Matic network, which offers a higher level of transparency of off-chain decision-making processes compared with traditional informal and unstructured off-chain governance [8]. Our proposed off-chain governance approach is also embedded with the multisig smart contract function to ensure transparent consensus vis-a-vis transaction information. The multisig smart contract built on the Polygon/Matic network not only identifies the users who are involved in finalising transaction rules and policies but can for example also record the process of how users agree on which data need to be validated and stored on the STN blockchain. This improves the transparency of off-chain decision-making processes. Further, the off-chain governance approach interacts with on-chain governance that equips the on-chain governance with flexible capability to be responsive to unforeseen circumstances.

## IV. PROOF OF CONCEPT DESIGN AND IMPLEMENTATION

This section reports the design architecture of the Proof-of-Concept (PoC) and its implementation in an Australian beef supply chain. The on-chain governance mechanism is designed with the multisig smart contract user interface on the STN Blockchain that enables supply chain actors to set on-chain transaction rules. The off-chain governance mechanism is built with the multisig smart contract user interface on the Polygon/Matic network that allows supply chain members to finalise off-chain rules. While on-chain and off-chain PoCs are built on different blockchain networks, they are linked by MetaMask and designed with compatibility on the STN network.

### A. PoC for On-chain Governance

The proof of concept for on-chain governance was built to track data about cattle which were registered on farms and whose subsequent changes of state were measured with transactions confirmed via a multisig smart contract protocol to ensure data integrity at the entry point. Fig. 2 shows the developed user interface for implementing on-chain multisig protocol. The smart contract was built as multisig wallets, which allows users to propose and agree to data state updates. Users can use the interface to set contracts, add and remove signature group members as well as change signing requirements. This user interface also allows users to check approved and pending transactions.

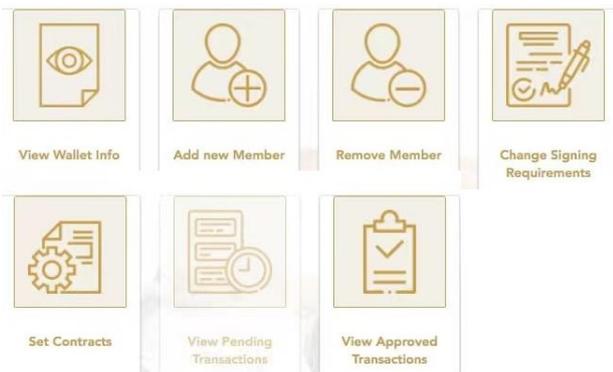

Fig. 2. User interface for on-chain multisig protocol

Fig. 3 gives a view of the data and cattle assets that are registered and measured by supply chain members within the system, which can offer a better understanding of how data is captured and presented in this PoC use case after applying the multisig smart contract protocol.

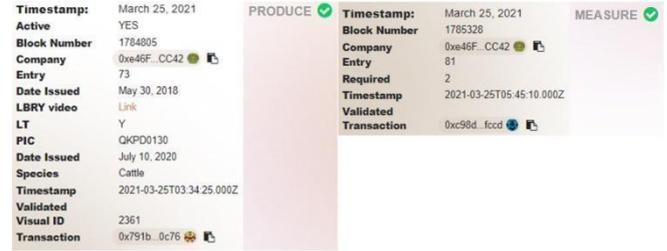

Fig. 3. Data and cattle asset views generated from on-chain multisig protocol

### B. PoC for Off-chain Governance

The proof of concept for off-chain governance was built to track decision-making processes in finalising rules and policies that are used to establish the protocols to govern on-chain transactions and off-chain storage. A simplified view of the designed user interface for implementing off-chain multisig protocol is shown in Fig. 3. The smart contract was built with STNBI Wallets on the Polygon/Matic network, which allows supply chain members to propose and agree on transaction governance rules. Supply chain members can use the interface to set multisig names, add/remove signers, change multisig requirements, transfer and mint tokens.

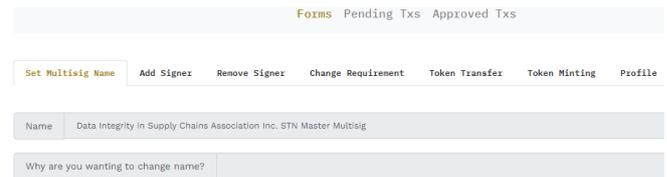

Fig. 4. User interface for off-chain multisig protocol

The application of our proposed off-chain governance on the Polygon /Matic network is demonstrated with an illustrative use case. Fig. 5 shows the history of decision-making processes in finalising governance rules and policies within the system. Guided by the multisig smart contract protocol, the decision-making process of off-chain governance rules and policies is recorded and disclosed with critical information, including destination, actions, participants, confirmation, and description of activities. It offers a higher level of transparency of off-chain decision-making processes in this PoC use case.

Fig. 5. User interface for off-chain multisig protocol

## V. Conclusion

This paper presents an integrated data governance framework that leverages multisig smart contracts in both on-chain and off-chain governance mechanisms to improve supply chain transparency and integrity from the data point of view. This paper furthers the conversation about the integration of on-chain and off-chain blockchain governance, especially the disclosure of off-chain decision-making processes on data governance, to strengthen supply chain transparency and integrity. We explore the applicability of our integrated data governance approach for practical implementation with the proof-of-concept design and implementation.

While our study is still work-in-progress, it makes both technical and theoretical contributions. The technical contributions include the development of multisig smart contracts and the development of a compatible network to accommodate two different blockchain networks for an integrated on-chain and off-chain governance approach. This paper also advances blockchain governance for supply chain transparency and integrity management from a theoretical point of view.

Our current design and proof-of-work bridges two different blockchain networks for on-chain and off-chain integration using MetaMask, which is one of the areas that need further investigation. Our further research will focus on cross-chain interoperability and design and evaluate cross-chain architecture that can enable more effective communications between two or more independent blockchain networks and consider the potential trade-off between efficiencies and security. The practical implementation with supply chain stakeholders will be included in the next stage of the project.

## Acknowledgment

The authors thank our study participants, project partners, and additional team members working on the larger Smart Trade Networks program of research, including Assoc. Prof. Felicity Deane and Santiago Del Valle. This project was supported by funding from the Future Food Systems CRC Ltd, funded under the Commonwealth Government CRC Program. The CRC Program supports industry-led collaborations between industry, researchers, and the community.